\documentclass[article,twocolumn,showpacs,aps,prb,floatfix,superscriptaddress,longbibliography]{revtex4-2}
\usepackage{graphicx,subfigure,epsfig,verbatim,psfrag,amsmath,amssymb,color}
\usepackage{indentfirst}
\usepackage{textcomp}
\usepackage{latexsym}
\usepackage{amssymb}
\usepackage{amsmath}
\usepackage{lipsum}
\usepackage{soul}
\usepackage{xcolor}
\usepackage{color}
\usepackage{epstopdf}
\usepackage{placeins}
\usepackage{float}

\usepackage{bm}

\newcommand {\R}{\textcolor {black}}

\usepackage{xcolor}
\usepackage[urlcolor=blue]{hyperref}      % links to citations
\hypersetup{
    colorlinks = true,                    % text and not border
    citecolor = {blue},
    linkcolor = {purple},
           }
\newcommand{\be}{\begin{equation}}
\newcommand{\ee}{\end{equation}}
\newcommand{\bea}{\begin{eqnarray}}
\newcommand{\eea}{\end{eqnarray}}

\begin{document}		
\title{Nature of the 1/9-magnetization plateau in the spin-1/2 kagome Heisenberg antiferromagnet}
\author{Da-zhi Fang}
%%\email{}
\affiliation{School of Physical Sciences, University of Chinese Academy of Sciences, Beijing 100049, China}
%\thanks{School of Physical Sciences, University of Chinese Academy of Sciences, P. O. Box 4588, Beijing 100049, China}
\author{Ning Xi}
%%\email{}
\affiliation{Department of Physics and Beijing Key Laboratory of Opto-electronic Functional Materials and Micro-Nano Devices, Renmin University of China, Beijing 100872, China}
\affiliation{CAS Key Laboratory of Theoretical Physics, Institute of Theoretical Physics, Chinese Academy of Sciences, Beijing 100190, China}
%%\date{}
%\thanks{School of Physical Sciences, University of Chinese Academy of Sciences, P. O. Box 4588, Beijing 100049, China}

\author{Shi-Ju Ran}\email[Corresponding author. ] {Email: sjran@cnu.edu.cn}
%\affiliation{Department of Physics, Capital Normal University, Beijing 10048, China}
\affiliation{Center for Quantum Physics and Intelligent Sciences, Department of Physics, Capital Normal University, Beijing 100048, China}
%\thanks{Department of Physics, Capital Normal University, Beijing 100048, China}
\author{Gang Su}
\email[Corresponding author. ] {Email: gsu@ucas.ac.cn}
\affiliation{School of Physical Sciences, University of Chinese Academy of Sciences, Beijing 100049, China}
\affiliation{Kavli Institute for Theoretical Sciences, and CAS Center for Excellence in Topological Quantum Computation, University of Chinese Academy of Sciences, Beijing 100190, China}
%\thanks{Kavli Institute for Theoretical Sciences, and CAS Center for Excellence in Topological Quantum Computation, University of Chinese Academy of Sciences, Beijing 100190, China}
%\thanks{School of Physical Sciences, University of Chinese Academy of Sciences, P. O. Box 4588, Beijing 100049, China}
\date{\today}

\begin{abstract}
    The nature of the 1/9-magnetization plateau of the spin-1/2 kagome Heisenberg antiferromagnet remains controversial due to the exotic physical properties and high complexity induced by the geometrical frustration. Instead of a Z$_{3}$ quantum spin liquid revealed on a cylinder, we show on the infinite-size lattice that the 1/9-plateau can be described by a valence bond crystal (VBC) that breaks the spatial translational invariance. Consistent results are achieved by two accurate tensor network methods, namely the full-update infinite projected-entangled pair states and the projected-entangled simplex states. The VBC exhibits an hourglass pattern with the $\sqrt{3}\times\sqrt{3}$ spatial symmetry, demonstrated by the magnetizations, the bond energies, and the three-body correlators. The spatial inversion symmetry in the $\sqrt{3}\times\sqrt{3}$ VBC is instantly broken with the presence of the difference between the coupling strengths in the up and down triangles, suggesting the existence of the gapless excitations. The gapless nature of the 1/9-plateau is further indicated by the scaling behaviors of the \R{entanglement entropy and the correlation length, which indicate a $c=1$ conformal field theory.}
\end{abstract}
\maketitle

%\section{Introduction}
\paragraph*{Introduction.---}

With the extensive interests in the novel physics induced by geometrical frustration, the ground state of the spin-1/2 kagome Heisenberg antiferromagnet (KHA) in an external magnetic field has been in a hot and long-lasting debate. Benefitted from the fast development on the efficient quantum many-body methods particularly those based on tensor networks (TN)~\cite{tn1qsl,tn2qsl,tn3qsl,tn4qsl,tn5qsl,tn6qsl,tn1,tnbook}, a series of exotic quantum states beyond the Landau-Ginzburg paradigm have been unveiled, such as valence bond crystals (VBC)~\cite{vbc1symh,vbc2symh,vbc3,vbc4,vbc5,ed1vbch,ed2vbch,ed3vbc,mc1vbc} and quantum spin liquids (QSL)~\cite{qsl1,ed1qsl,mc1qsl,mc2qsl,mc3qsl,mc4qsl,mc5qsl,mc6qsl,dmrg1qsl,dmrg2qsl,dmrg3qsl,dmrg4qsl,dmrg5qsl,dmrg6qsl,dmrg7qsl,tn1qsl,tn2qsl,tn3qsl,tn4qsl,tn5qsl,tn6qsl}, etc. 

Among others, previous works using the density matrix renormalization group (DMRG) uncovered that an external magnetic field can induce an unconventional 1/9-magnetization plateau in a finite-size KHA cylinder, which might be a topological Z$_{3}$ QSL~\cite{dmrg4qsl}. This is exotic since the external fields usually tend to suppress the topological orders and give a conventional phase. In contrast, the magnetic ordering for the 1/3, 5/9 and 7/9 plateaus in this system can be well described by the ordinary magnetic structures~\cite{vbc1symh,vbc2symh,ed1vbch,ed2vbch,dmrg4qsl}. However, the nature of the 1/9-plateau remains controversial due to the limitations of the DMRG algorithm. It is unclear how significantly the finite-size effects could affect the physics of the 1/9-plateau, and whether the Z$_{3}$ QSL can survive in the thermodynamic limit.

In this work, we apply the infinite projected-entangled pair states (iPEPS)~\cite{ipeps1full,ipeps2full,ipeps3full} and the 9-projected-entangled simplex states (9-PESS)~\cite{tn1} to study the 1/9-magnetization plateau of the spin-1/2 KHA. Both algorithms are developed for simulating infinite-size systems and accurately give the magnetic fields for the 1/9-plateau, which coincides with those previously obtained by the cluster update of the iPEPS~\cite{tn4qsl}. The magnetizations, the bond energies, and the three-body correlators consistently show that the 1/9-plateau should be described by a VBC instead of a Z$_{3}$ QSL. The VBC breaks the spatial translational invariance and forms a special hourglass configuration satisfying the $\sqrt{3}\times\sqrt{3}$ symmetry~\cite{vbc1symh,vbc2symh,ed8symh,ed10symh,ipeps3cluster}. Besides, we show that the spatial inversion symmetry in the $\sqrt{3}\times\sqrt{3}$ VBC is not protected by any gap. The presence of the difference between the coupling strengths inside the up and down triangles will instantly break this symmetry. The gapless nature of the 1/9-plateau can be further revealed by the dominant eigenstate of the transfer matrix of the iPEPS. \R{We demonstrate the logarithmic scaling of the entanglement entropy and the algebraic scaling of the correlation length against the virtual bond dimension of the boundary matrix product states (MPS)~\cite{37}. Both scaling behaviors show that the VBC should be described by a $c=1$ conformal field theory (CFT)~\cite{CFTbook,CFT,ccfbf1,ccfbf2}.}
\\

\paragraph*{Model and Methods.---}

\begin{figure*}[tbp]
	\includegraphics[width=0.9\linewidth]{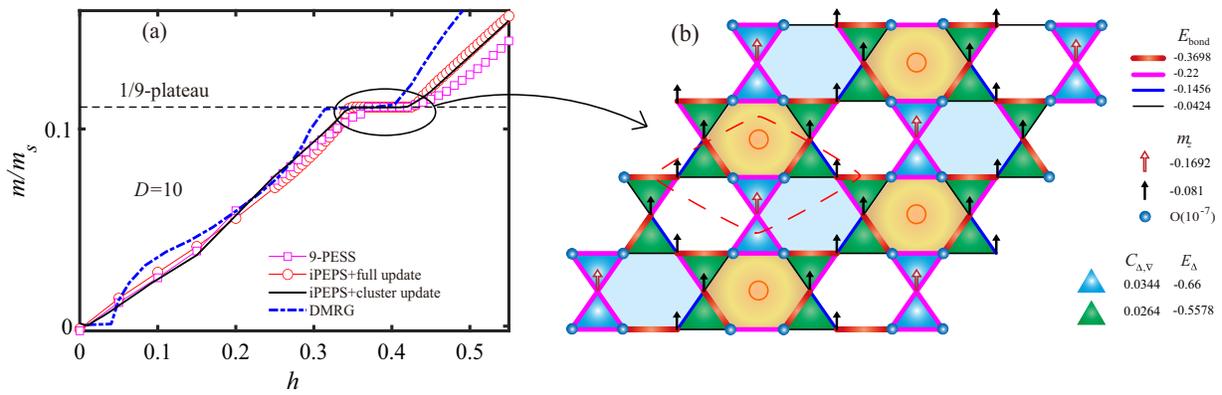}
	\caption{(a) The average magnetization per site $m/m_{s}$ versus the external field $h$. We fix the virtual bond dimension of the iPEPS to be $D=10$. The magnetizations obtained by the DMRG~\cite{dmrg4qsl} (blue dash line) and the cluster update of the iPEPS~\cite{tn4qsl} (black solid line) are also given for comparison. In (b) we show the local magnetization in the spin-$z$ direction $m_{z}$, the bond energies $E_{\rm bond}$, the energies of triangles $E_{\Delta}$, and the three-body correlators $C_{\Delta, \nabla}$ in the 1/9-plateau phase. These quantities consistently show a special hourglass $\sqrt{3}\times\sqrt{3}$ structure, breaking the spatial translational invariance with an extended nine-site unit cell. Each orange circle in the hexagon represents four blue balls and two black arrows around it. The pattern shows that the hourglasses (filled with cyan and with red arrows in the center) form the triangular lattice. The red dash diamond, containing the orange circle, the red arrow, and two black arrows, shows the unit cell.}
\label{1}
\end{figure*}

\begin{figure*}[tbp]
	\includegraphics[width=0.9\linewidth]{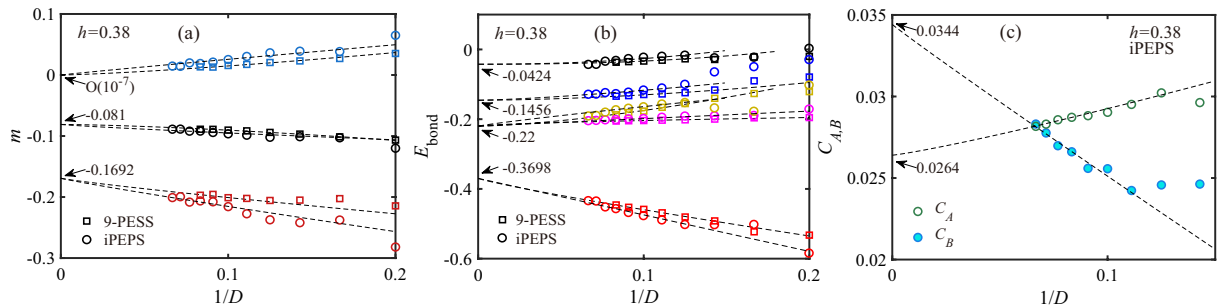}
	\caption{\R{(color online)} (a) The magnetizations $m$, (b) the bond energies $E_{\rm bond}$ and (c) the three-body correlators $C_{A,B}$ versus the virtual bond dimension $1/D$ of the iPEPS or the 9-PESS. The magnetic field is chosen to be on the 1/9-plateau with $h=0.38$. The dash lines give the extrapolations towards the $1/D\to0$ limit. \R{In (b), we find that the extrapolation of the bond energy colored by orange is consistent with the pink one. Therefore, in Fig.~\ref{1}(b) we uniformly use pink to represent these kinds of bond energies. We finally have four kinds of bond energies, corresponding to those shown in Fig.~\ref{1}(b). We take the virtual bond dimension of the boundary MPS $\chi=25$ in these simulations.}}
\label{2}
\end{figure*}

Consider the spin-1/2 KHA in a magnetic field, and the Hamiltonian reads
\begin{equation}
\hat{H}=\sum_{\langle i,j\rangle} \bm{\hat{S_{i}}} \cdot \bm{\hat{S_{j}}}-h\sum_{i}\hat{S^{z}_{i}},
\end{equation}
with $\bm{\hat{S_{i}}}$ the spin operator on the $i$th site, $\hat{S^{z}_{i}}$ the $z$ component of $\bm{\hat{S_{i}}}$, $h$ the external field along the $z$ direction, and $\langle i,j\rangle$ two nearest-neighbor spins.

We here apply two accurate TN methods, namely the iPEPS~\cite{ipeps1full,ipeps2full,ipeps3full} and the 9-PESS~\cite{tn1}, to implement our numeric simulations. In particular, for the iPEPS algorithm, we use the full update~\cite{ipeps1full,ipeps2full,ipeps3full} scheme to optimize the local tensors. The TN contractions in the full update are calculated by the infinite time-evolving block decimation (iTEBD)~\cite{itebd}. The full update scheme, though more expensive than the simple~\cite{ipeps1simple} and the cluster~\cite{ipeps1cluster,ipeps2cluster,ipeps3cluster} update schemes, allows to implement the optimizations and simulate the observables more accurately by better considering the quantum correlations in the infinite-size system. The details of these two algorithms can be found in the supplemental material~\cite{SM} (see also references~\cite{tn3qsl,tn1,ipeps1full,ipeps2full,ipeps3full,itebd,TD,ctmrg} therein).
\\

\paragraph*{Nature of the 1/9-magnetization plateau.---}

Let us start from the ground-state phase diagram in an external magnetic field. Fig.~\ref{1}(a) shows the average magnetization per site $m/m_{s}$ versus the magnetic field $h$. The $m/m_{s}=1/9$ magnetization plateau (with $m_{s}=1/2$ the saturated magnetization) appears in the range of $0.35<h<0.42$ from both the iPEPS and the 9-PESS. This is also qualitatively consistent with previous works by the cluster update of the iPEPS~\cite{tn4qsl} (solid line) and the DMRG on a finite-size cylinder~\cite{dmrg4qsl} (dash line). Note for $h=0$, the DMRG gives a gapped ground state with a $m=0$ zero plateau, while the rest three methods show a gapless ground state whose the magnetization immediately becomes non-zero by adding an external field. It is likely that the gap leading to the zero plateau is a result of the finite-size effects.

\begin{figure}[tbp]
	\includegraphics[scale=0.4]{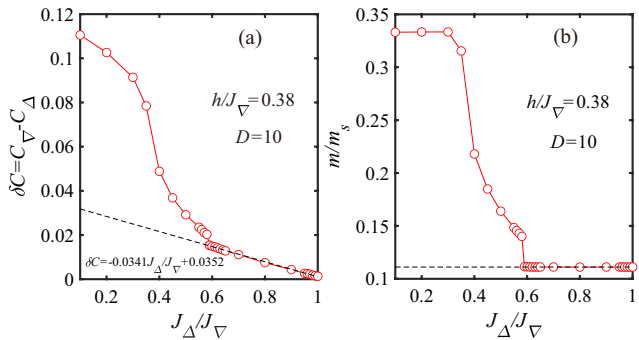}
	\caption{(a) The difference of three-body correlators $\delta C=C_{\nabla}-C_{\Delta}$, and (b) the average magnetization per site $m/m_{s}$ versus $J_{\Delta}/J_{\nabla}$ of the breathing kagome model with $h/J_{\nabla}=0.38$. We still fix the virtual bond dimension of the iPEPS to be $D=10$. In the range of $0.59<J_{\Delta}/J_{\nabla}<1$, $m/m_{s}$ remains in the 1/9-plateau. The width of the plateau indicates the gap that protects the $\sqrt{3}\times\sqrt{3}$ phase against the magnetic field. In the same region of $J_{\Delta}/J_{\nabla}$, $\delta C$ grows linearly as soons as $J_{\Delta}/J_{\nabla}$ becomes non-zero.}
\label{3}
\end{figure}

Focusing on the main target of this work, which is the $1/9$-magnetization plateau, Fig.~\ref{1}(b) shows the local magnetization in the spin-$z$ direction $m_{z}$, the bond energies $E_{\rm bond}$, the energies of triangles $E_{\Delta}$, and the three-body correlators $C_{\Delta, \nabla}$ (see the definitions and calculations of these quantities in the supplemental material~\cite{SM}). We take $h=0.38$ within the 1/9-plateau without losing generality. Fig.~\ref{2}(a) and (b) show the differences of these quantities between the iPEPS and the 9-PESS, which are about $10^{-2}$ for $D=5$ and vanish as $D$ increases. \R{Our results show that the observables converge fast with the virtual bond dimension $\chi$ of the boundary MPS, where the changes become insignificant for about $\chi>10$.}

By extrapolation, we identify three kinds of magnetizations and four bond energies, marked by different arrows and lines in Fig.~\ref{1}(b). The spins with vanishing magnetizations ($m\sim O(10^{-7})$) are marked by blue balls. An hourglass pattern exhibiting the $\sqrt{3}\times\sqrt{3}$ symmetry is uncovered. There are two kinds of three-body correlators, defined by $C_{A,B}$ shown in Fig.~\ref{2}(c). \R{Three types of hexagons can be defined according to these quantities, which are colored by orange, white, and sky blue. These hexagons appear periodically, forming a $\sqrt{3}\times\sqrt{3}$ pattern on a triangular lattice~\cite{ed8symh,ipeps3cluster}. The unit cell is illustrated by the red dash diamond.}

\begin{figure}[htb]
	\includegraphics[scale=0.4]{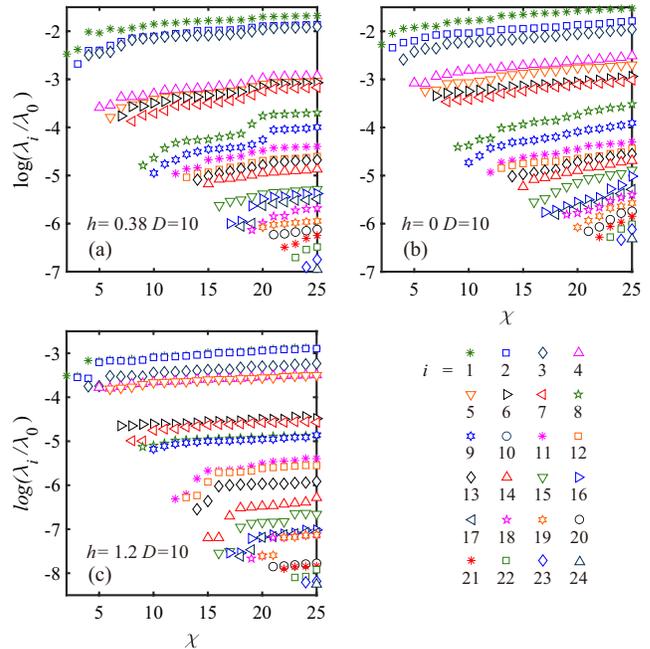}
	\caption{The entanglement spectra $\rm log(\lambda_{i}/\lambda_{0})$ of the transfer matrix of the iPEPS on the spin-1/2 KHA versus the virtual bond dimension $\chi$ of the boundary MPS. We fix the virtual bond dimension of the iPEPS to be $D=10$. The magnetic field is chosen to be (a) on the 1/9-plateau with $h=0.38$, (b) $h=0$ a gapless QSL, and (c) on the 1/3-plateau with $h=1.2$.}
\label{4}
\end{figure}

In a conventional magnetic plateau phase, there usually exists a finite excitation gap characterized by the width of the plateau. We consider the breathing kagome model by changing the coupling strengths inside the up and down triangles, whose Hamiltonian can be written as
\begin{equation}
\hat{H}=J_{\nabla}\sum_{\langle i,j\rangle}\bm{\hat{S_{i}}}\cdot \bm{\hat{S_{j}}}+J_{\Delta}\sum_{\langle i,j\rangle}\bm{\hat{S_{i}}}\cdot \bm{\hat{S_{j}}}-h\sum_{i}\hat{S^{z}_{i}},
\end{equation}
with $J_{\Delta}$ and $J_{\nabla}$ the coupling strengths in the up and down triangles. Fig.~\ref{3} shows the average magnetization per site $m/m_{s}$ and the difference of the three-body correlators $\delta C=C_{\nabla}-C_{\Delta}$ versus $J_{\Delta}/J_{\nabla}$.

We find that $m/m_{s}$ remains in the 1/9-magnetization plateau in the range of $0.59<J_{\Delta}/J_{\nabla}<1$, indicating the gap that protects the $\sqrt{3}\times\sqrt{3}$ phase against the magnetic field. $\delta C$ grows linearly when $J_{\Delta}/J_{\nabla}$ becomes non-zero in the same region, which shows that the spatial inversion symmetry in the $\sqrt{3}\times\sqrt{3}$ VBC will instantly be broken once the coupling strengths in the up and down triangles have a difference. These quantities suggest the existence of the gapless non-magnetic excitations. It gives us an evidence that the 1/9-plateau with $J_{\Delta}=J_{\nabla}$ is a gapless critical system.

\begin{figure}[htb]
	\includegraphics[scale=0.4]{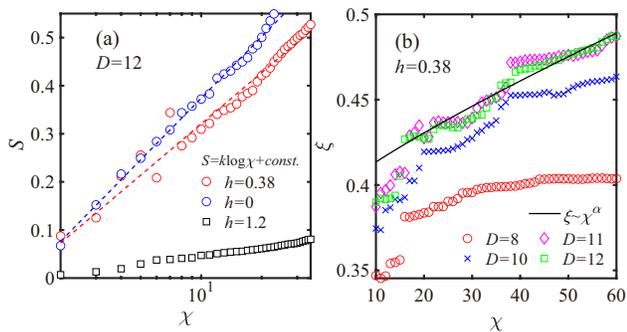}
	\caption{\R{(a) The entanglement entropy $S$ versus $\chi$ (log-scale). We fix the virtual bond dimension of the iPEPS to be D = 12.The dash lines give the logarithmic fitting of $h=0.38$ with $k\simeq0.154$, and $h=0$ with $k\simeq0.185$. (b) The correlation length $\xi$ versus $\chi$ with different virtual bond dimensions of the iPEPS. The black line gives the power fitting with $\alpha\simeq0.841$.}}
\label{5}
\end{figure}

\R{In the critical or gapped systems, the correlation length, the entanglement spectrum, and the entanglement entropy exhibit different scaling behaviors against the virtual bond dimension $\chi$ of the boundary MPS~\cite{37,scaling1,scaling2,scaling3}. Fig.~\ref{4} shows the logarithmic entanglement spectra $\rm log(\lambda_{i}/\lambda_{0})$ of the boundary MPS by varying the virtual bond dimension $\chi$. For $h=0.38$ (on the 1/9-plateau), we see in Fig.~\ref{4}(a) that the elements of the entanglement spectrum are squeezed as $\chi$ increases~\cite{37}. This result is consistent with the case for $h=0$ shown in Fig.~\ref{4}(b), where a gapless QSL was revealed~\cite{tn3qsl,tn4qsl,tn5qsl,tn6qsl,mc5qsl,dmrg5qsl}. In comparison, the elements of the spectrum stay almost unchanged for $h=1.2$ (gapped 1/3-plateau), as shown in Fig.~\ref{4}(c). These results indicate that the 1/9-plateau should be a gapless phase.}

\R{To further uncover the gapless nature of the 1/9-plateau, we calculate the entanglement entropy $S$ and the correlation length $\xi$ of the boundary MPS, which are defined as
\begin{eqnarray}
S&=&-\sum_{i}\lambda_{i}^{2}\rm log\lambda_{i}^{2}, \\
\xi&=&\frac{1}{\rm log |\frac{\Lambda_0}{\Lambda_1}|},
\end{eqnarray}
with $\Lambda_i$ the $i$th eigenvalue of the transfer matrix of the boundary MPS~\cite{37,c1MPS}. Fig.~\ref{5} shows the scalings of $S$ and $\xi$ against $\chi$. Obvious logarithmic scalings of $S$ are demonstrated in Fig.~\ref{5}(a) for $h=0$ and $h=0.38$ (1/9-plateau), consistent with the results in Fig.~\ref{4}. For $h=0.38$, the algebraic scaling of $\xi$ converges for $D \geq 11$ [Fig.~\ref{5}(b)]. Critical scaling laws are obtained, obeying
\begin{eqnarray}
S&=&k\rm log\chi+const., \\ 
\xi &\sim& \chi^\alpha,
\end{eqnarray}
with $k\simeq0.154$ and $\alpha\simeq0.841$ by fitting the data of $D=12$ and $\chi \geq 20$. The scalings of $S$ and $\xi$ allow to estimate the central charge $c$ as~\cite{CFTbook,CFT}
\begin{equation}
c=\frac{6k}{\alpha}.
\end{equation}
We have $c=1 + O(10^{-1}) \simeq1$. The fluctuation is caused by the finiteness of the virtual bond dimensions of the iPEPS and the boundary MPS.}

\R{From CFT~\cite{CFTbook,CFT,ccfbf1,ccfbf2}, the central charge $c$ indicates the degrees of freedom of a critical system. Previous works on the spin-1/2 KHA suggested that the ground state with zero magnetic field should be a gapless QSL, which can be reduced to a $c=1$ Luttinger liquid by the U(1) gauge field~\cite{dmrg5qsl}. For the J$_1$-J$_2$-J$_3$ model of the spin-1/2 KHA, the magnetically-ordered phase, known as ``cuboc1''~\cite{50}, is found to consist of three gapless free bosonic modes with $c=3$ by fully breaking the SU(2) symmetry~\cite{cc1}. Our results show that the 1/9-plateau is a gapless phase with $c=1$, where the $\sqrt{3}\times\sqrt{3}$ VBC could be described as a free bosonic field by breaking the spatial inversion symmetry~\cite{ccfbf1,ccfbf2}.}
\\

\paragraph*{Conclusion.---}

In this work, we employ the iPEPS with full update and the 9-PESS methods to study the nature of the 1/9-magnetization plateau of the spin-1/2 KHA in an external magnetic field. The 1/9-plateau is found in the range of $0.35<h<0.42$, consistent with previous works by the cluster update of the iPEPS and the DMRG. We do not observe the zero plateau as $h\to0$, which is different from a series of DMRG simulation on the finite-size systems~\cite{dmrg2qsl,dmrg3qsl,dmrg4qsl}. 

We calculate the magnetizations $m_{z}$, the bond energies $E_{\rm bond}$ and the three-body correlators $C_{\Delta,\nabla}$ on the 1/9-plateau. By the extrapolations towards the $1/D\to0$ limit, these results consistently show that the 1/9-plateau should be described by a VBC that breaks the spatial translational invariance instead of a Z$_{3}$ QSL. The VBC exhibits a special hourglass pattern satisfying the $\sqrt{3}\times\sqrt{3}$ symmetry. Introducing a difference between the coupling strengths in the up and down triangles, the spatial inversion symmetry in the $\sqrt{3}\times\sqrt{3}$ VBC is simultaneously broken. \R{The gapless nature of the VBC is also revealed by the logarithmic scaling behavior of the entanglement entropy $S$, and the algebraic scaling behavior of the correlation length $\xi$. By fitting, the central charge $c$ implies that the VBC could be described by a $c=1$ free bosonic field.}

\begin{acknowledgments}
This work is supported in part by the NNSFC (Grant No.~11834014 and No.~12004266), the Strategic Priority Research Program of CAS (Grant No.~XDB280000000), the National Key R\&D Program of China (Grant No.~2018YFA0305800), Beijing Natural Science Foundation (Grant No.~1232025), R\&D Program of Beijing Municipal Education Commission (No.~KM202010028013), and Academy for Multidisciplinary Studies, Capital Normal University.
\end{acknowledgments}	

%

%\input{artical.bbl}
%\bibliographystyle{unsrt}
%\bibliography{ref_prb}
\end{document}